\documentclass[review]{elsarticle}

\usepackage{hyperref}
\usepackage[mathlines]{lineno}
\modulolinenumbers[1]

\journal{Journal of Computational Physics Templates}

\usepackage{amsmath, graphicx}
\usepackage{grffile}
\usepackage{dcolumn}
\usepackage{bm}
\usepackage{url}
\usepackage{epsfig}
\usepackage{mathrsfs}  
\usepackage{subcaption}
\usepackage{multirow}
\usepackage{algorithm}
\usepackage{algorithmicx}
\usepackage{amssymb}
\usepackage{tensor}
\usepackage{romannum}
\usepackage{listings}

\usepackage{amsfonts}
\usepackage{amssymb}
\usepackage{graphicx}
\usepackage{algpseudocode}
\usepackage{algorithm}
\usepackage{algorithmicx}
\usepackage{xcolor}
\usepackage{fancyhdr}
\usepackage{subfiles}
\usepackage[makeroom]{cancel}
\usepackage{algpseudocode}

\usepackage{hyperref}
\hypersetup{
    colorlinks=true,
    linkcolor=blue,
    filecolor=magenta,      
    urlcolor=cyan,
}

\usepackage{forest}

\definecolor{folderbg}{RGB}{124,166,198}
\definecolor{folderborder}{RGB}{110,144,169}

\def\Size{4pt}
\tikzset{
  folder/.pic={
    \filldraw[draw=folderborder,top color=folderbg!50,bottom color=folderbg]
      (-1.05*\Size,0.2\Size+5pt) rectangle ++(.75*\Size,-0.2\Size-5pt);  
    \filldraw[draw=folderborder,top color=folderbg!50,bottom color=folderbg]
      (-1.15*\Size,-\Size) rectangle (1.15*\Size,\Size);
  }
}

\graphicspath{{Figures/}}

\begin{document}

\begin{frontmatter}

\title{A Spacetime Finite Elements Method to Solve the Dirac Equation}

\author{Rylee Sundermann$^{1}$, Hyun Lim$^{2,3,4}$, Jace Waybright $^{6,1}$, Jung-Han Kimn$^{1}$}
\address{1. Department of Mathematics and Statistics, South Dakota State University, Brookings, SD 57007 USA}
\address{2. The Computational Physics and Methods Group, Los Alamos National Laboratory, Los Alamos, NM 87545 USA}
\address{3. Center for Theoretical Astrophysics, Los Alamos National Laboratory, Los Alamos, NM 87545 USA}
\address{4. Applied Computer Science Group, Los Alamos National Laboratory, Los Alamos, NM 87545 USA}
\address{5. Princeton Plasma Physics Laboratory, Princeton University, Princeton, New Jersey 08543, USA}


\begin{abstract}
In this work, a fully implicit numerical approach based on
space-time finite element method is presented to solve the Dirac equation in 1 (space) + 1 (time), 2 + 1, and 3 + 1 dimensions. We utilize PETSc/Tao library to implement our linear system and for using Krylov subspace based solvers such as GMRES. We demonstrate our method by analyzing several different cases including plane wave solution, Zitterbewegung, and Klein paradox. Parallel performance of this implementation is also presented.
\end{abstract}

\begin{keyword}
\texttt{Finite Elements}\sep \texttt{Dirac Equation} \sep \texttt{PETSc} 
\end{keyword}

\end{frontmatter}


\section{Introduction}
The Dirac equation governs spin 1/2 particles, known as fermions. It has been applied
and studied extensively to many fields of physics and chemistry including
relativistic heavy ion collisions, heavy ion spectroscopy, laser-matter interaction,
and condensed matter physics~\cite{Salamin2006,Katsnelson2006}.
Although the Dirac equation has been utilized profusely, solving this equation is still a 
challenging problem. Due to the complicated nature of the Dirac equation, only highly
symmetric systems can be studied analytically, i.e. the more realistic case should be
based on an approximation-based method such as semi-classical theory~\cite{Milosevic2002} 
and numerical calculations. However, the usual time scales of the fermion dynamics 
is often much smaller than the time scales of interesting phenomena. Thus, obtaining
the numerical solutions is extremely difficult, and it is hard to maintain computational
efficiency. Furthermore, certain numerical schemes, such as naive symmetric spatial differencing,
are often encountered with fermion doubling problem~\cite{Muller1998,Richard1982,Kogut1975}. 
Previously, there are several different numerical approaches to explore the Dirac equation. 
In~\cite{Braun1999,Mocken2008,Mocken2004,Bauke2011,Momberger1996}, some variations of the
operator splitting method along with a spectral scheme are presented. Also, the finite element scheme~\cite{Muller1998,Bottcher1985}
and finite difference scheme~\cite{Becker1983,Selsto2009,Gelis2005} have been exploited.

In this work, we apply spacetime finite element method (FEM) to solve the Dirac equation 
in 1 (space) + 1 (time), 2 + 1, 3 + 1 dimensions. 
This work applies spacetime FEM in a similar way as in previous works~\cite{AndersonJCP2007,LimSURIO2013,Lim2014}. 
The spacetime FEM has advantages for numerical simulations. The method explored
in this work is a fully-implicit method. Instead of time integration, spatial and temporal dimensions are considered simultaneously. Thus, approximated solutions can avoid accumulated time integration errors. Furthermore, this approach can use time varying computational domain, higher order approaches, and unstructured meshes. 
Moderate size of simulation based on the spacetime FEM requires solving linear systems with millions of unknowns. Therefore, the major challenge of the method is the significant memory overhead requirement that entire spacetime problem needs to fit in memory all at once. 

The rest of this paper is organized as follows: in Section~\ref{sec:numerical}, 
the numerical approach is provided, including space-time finite element discretization 
and solvers, in Section~\ref{sec:software}, implementation description of 
our work is presented;
Section~\ref{sec:results} presents results with different example cases while Section~\ref{sec:conclusion} 
contains the conclusions and future works

\section{Numerical Approaches}
\label{sec:numerical}
\subsection{Weak Formulation of the Dirac Equation}
\label{sec:sub:weakDirac}
The standard form of the Dirac equation in a gauge-free case is:
\begin{eqnarray}
\label{eqn:dirac}
(i\hbar\gamma^\mu\partial_\mu - m)\Psi(x^\mu) = 0
\end{eqnarray}
where $x^\mu \in \mathbb{R}^4$, a four dimensional Minkowski space. Traditionally $\mu = 0$ corresponds to time and $\mu = \,1, \, 2, \, 3$ corresponds to the three spatial dimensions. In this work, we adapt the natural unit system such as $c = \hbar = 1$.

And, the weak form of the Dirac boundary value problem is to find $\Psi_h$ such that:
\begin{eqnarray}
\int_\Omega \overline\Phi_h(i\hbar\gamma^\mu\partial_\mu - m)\Psi_h d\Omega = 0 \nonumber \\
\Psi_h(x^i) = \Psi_0(x^i) \text{ for } x^i \in \partial \Omega
\end{eqnarray}

for every $\Phi_h$ where:
\begin{eqnarray}
\Phi_h = \begin{bmatrix}  \Phi_1 n_1(x^\mu) & \Phi_2 n_2(x^\mu) & ... & \Phi_N n_N(x^\mu) \end{bmatrix}^T \nonumber \\
\Psi_h = \begin{bmatrix}  \Psi_1 n_1(x^\mu) & \Psi_2 n_2(x^\mu) & ... & \Psi_N n_N(x^\mu) \end{bmatrix}^T \nonumber \\
N \in \mathbb N, \Phi_i, \Psi_i \in \{ \begin{bmatrix} \phi_1 & \phi_2  & \phi_3 & \phi_4 \end{bmatrix} ^T \phi_{1-4} \in \mathbb C\} \nonumber \\
n(x^\mu) : \mathbb{R}^4 \rightarrow \mathbb{R}\nonumber 
\end{eqnarray}
where $\mathbb R^4$ is the 4D Minkowski space, and $\overline\Phi_h = \Phi^* \gamma_0$. 
Note that both $\Phi_h$ and $\Psi_h$ share the same set of interpolation functions $n(x^\mu)$.
Substituting for the definition of $\Phi_h$ and $\Psi_h$ allows us to move to the matrix form and simplify the expression.
\begin{eqnarray}
\int_\Omega \begin{bmatrix}  \overline\Phi_1 n_1(x^\mu) & \overline\Phi_2 n_2(x^\mu) & ... & \overline\Phi_N n_N(x^\mu) \end{bmatrix}  (i\hbar\gamma^\mu\partial_\mu - m) \begin{bmatrix}  \Psi_1 n_1(x^\mu) \\ \Psi_2 n_2(x^\mu)\\ ... \\ \Psi_N n_N(x^\mu) \end{bmatrix}  d\Omega = 0 \
\end{eqnarray}
The function vectors $\Phi_h$ and $\Psi_h$ can be simplified
by using the subscripts $i$ and $j$ to refer to 
the columns and rows of the resulting matrix. Thus, we can
rewrite this as follows
\begin{eqnarray}
\left[ \int_\Omega \overline\Phi_j n_j(x^\mu) (i\hbar\gamma^\mu\partial_\mu - m)  \Psi_i n_i(x^\mu) d\Omega  \right]_{ ij} = 0 \;
\text{ where } \; 1\leq i,j \leq N
\end{eqnarray}
All of the complex constants in $\Phi_h$ can then be factored out 
to the front of the matrix
\begin{eqnarray}
\overline\Phi \left[ \int_\Omega n_j(x^\mu) (i\hbar\gamma^\mu\partial_\mu - m)   n_i(x^\mu) d\Omega  \right]_{ij}\Psi_i = 0 \;\;
\text{where} \;\;  \overline\Phi = \begin{bmatrix}  \overline\Phi_1 & \overline\Phi_2  & ... & \overline\Phi_N  \end{bmatrix}
\end{eqnarray}
and eliminated by multiplying both sides by their inverse
\begin{eqnarray}
\frac{\Phi}{\|\Phi\|^2}\cdot \overline\Phi \left[ \int_\Omega n_j(x^\mu) (i\hbar\gamma^\mu\partial_\mu - m) n_i(x^\mu) d\Omega  \right]_{ ij}\Psi_i  = \frac{\Phi}{\|\Phi\|^2}\cdot 0\nonumber  \\
\Rightarrow \left[ \int_\Omega n_j(x^\mu) (i\hbar\gamma^\mu\partial_\mu - m) n_i(x^\mu) d\Omega  \right]_{ ij}\Psi_i  = 0 \label{eqn:diracweak}
\end{eqnarray}

We will refer to Eqn.~\ref{eqn:diracweak} as the simplified weak form written as
\begin{eqnarray}
\label{eqn:Dw}
D_w \Psi = 0 
\end{eqnarray}

Here $D_w$ is a $N\times N$ block matrix with each entry a $4\times 4$ matrix giving a total dimension of $4N \times 4N$.
We begin by supposing that the test functions are chosen such that the boundary value at $t=0$ defines the first $K\in\mathbb N$ spinor values of $\Psi_h$ uniquely, such that $\Psi_0(\mathbf x^\mu) =  \begin{bmatrix}  \Psi_1 n_1(x^\mu) & \Psi_2 n_2(x^\mu)& ... & \Psi_K n_K(x^\mu) \end{bmatrix}$. This allows us to partition $D_w \Psi$ into known and unknown components as follows.
\begin{eqnarray}
\begin{bmatrix} 
\begin{bmatrix} 
d_{1,1} & \cdots & d_{1,K} \\ 
\vdots & D_{w11} & \vdots \\
d_{K,1} & \cdots & d_{K,K}
\end{bmatrix} & 
\begin{bmatrix} 
d_{1,K+1} & \cdots & d_{1,N} \\ 
\vdots & D_{w12} & \vdots \\
d_{K,K+1} & \cdots & d_{K,N}
\end{bmatrix} \\
\begin{bmatrix} 
d_{K+1,1} & \cdots & d_{K+1,K} \\ 
\vdots & D_{w21} & \vdots \\
d_{N,1} & \cdots & d_{N,K}
\end{bmatrix} & 
\begin{bmatrix} 
d_{K+1,K+1} & \cdots & d_{K+1,N} \\ 
\vdots & D_{w22} & \vdots \\
d_{N,K+1} & \cdots & d_{N,N}
\end{bmatrix}  
\end{bmatrix}
\times
\begin{bmatrix} 
\begin{bmatrix} 
\Psi_1 \\ \vdots \\ \Psi_K 
\end{bmatrix} \\ 
\begin{bmatrix} 
\Psi_{K+1} \\ \vdots \\ \Psi_{N} 
\end{bmatrix} 
\end{bmatrix} = 0
\end{eqnarray}
Using block matrix multiplication, this can be rewritten as 
the sum of four smaller matrix operation as follows.
\begin{eqnarray}
\begin{bmatrix} 
d_{1,1} & \cdots & d_{1,K} \\ 
\vdots & D_{w11} & \vdots \\
d_{K,1} & \cdots & d_{K,K}
\end{bmatrix} 
\begin{bmatrix} 
\Psi_1 \\ \vdots \\ \Psi_K 
\end{bmatrix}
+
\begin{bmatrix} 
d_{1,K+1} & \cdots & d_{1,N} \\ 
\vdots & D_{w12} & \vdots \\
d_{K,K+1} & \cdots & d_{K,N}
\end{bmatrix}
\begin{bmatrix} 
\Psi_{K+1} \\ \vdots \\ \Psi_{N} 
\end{bmatrix} 
+  ... \nonumber \\
\begin{bmatrix} 
d_{K+1,1} & \cdots & d_{K+1,K} \\ 
\vdots & D_{w21} & \vdots \\
d_{N,1} & \cdots & d_{N,K}
\end{bmatrix} 
\begin{bmatrix} 
\Psi_1 \\ \vdots \\ \Psi_K 
\end{bmatrix}
+
\begin{bmatrix} 
d_{K+1,K+1} & \cdots & d_{K+1,N} \\ 
\vdots & D_{w22} & \vdots \\
d_{N,K+1} & \cdots & d_{N,N}
\end{bmatrix}  
\begin{bmatrix} 
\Psi_{K+1} \\ \vdots \\ \Psi_{N} 
\end{bmatrix} 
= 0 
\end{eqnarray}
From the definition of the matrix, the first line corresponding to spinor
values $\Psi_{1 \rightarrow K}$ is equal to zero. Thus, we may remove it and
rewrite the equation as follow.
\begin{eqnarray}
\begin{bmatrix} 
d_{K+1,K+1} & \cdots & d_{K+1,N} \\ 
\vdots & D_{w22} & \vdots \\
d_{N,K+1} & \cdots & d_{N,N}
\end{bmatrix}  
\begin{bmatrix} 
\Psi_{K+1} \\ \vdots \\ \Psi_{N} 
\end{bmatrix} 
= 
-\begin{bmatrix} 
d_{K+1,1} & \cdots & d_{K+1,K} \\ 
\vdots & D_{w21} & \vdots \\
d_{N,1} & \cdots & d_{N,K}
\end{bmatrix} 
\begin{bmatrix} 
\Psi_1 \\ \vdots \\ \Psi_K 
\end{bmatrix}
\end{eqnarray}
Therefore, the equation can be written as using only the matrix subscript as
\begin{eqnarray}
\label{eqn:dirac_weakIVP}
D_{w22} \Psi_{K+1 \rightarrow N} = - D_{w21} \Psi_{1 \rightarrow K}
\end{eqnarray}

We also can express Eqn.~\ref{eqn:dirac_weakIVP} in terms of the boundary value problems
\begin{eqnarray}
\label{eqn:bp:dirac}
\left[ \int_\Omega n_j(x^\mu) (i\hbar\gamma^\mu\partial_\mu - m) n_i(x^\mu) d\Omega  \right]_{ij}\Psi_i 
=
-\left[ \int_\Omega n_k(x^\mu) (i\hbar\gamma^\mu\partial_\mu - m) n_i(x^\mu) d\Omega  \right]_{ik}  \Psi_k  \label{eqn:diracweakIVP} 
\\
\text{ where } K+1\leq i,j \leq N \text{ and } 1\leq k \leq K \nonumber
\end{eqnarray}

Equation~\ref{eqn:bp:dirac} is the weak formulation of the Dirac equation. Each $n_j$ is interpolate functions for finite element space.

\subsection{Space-Time Finite Element Discretization}
\label{sec:sub:spFEM}
A space-time FEM using continuous approximation functions in both space and time is used to solve the system. The space-time FEM divides the domain into a finite number of subsets, which are called elements, and confines the function space of the weak form to test functions that are non-zero on only a finite number of elements in the domain $\Omega$. Since the test functions have no value outside of the local element, we note that Eqn.~\ref{eqn:bp:dirac} is only non-zero when $n_i$ and $n_j$ belong to
the same finite element. Thus, we compute Eqn.~\ref{eqn:bp:dirac} as an integral over the basis function ($n_i$) of each element, which results in the element stiffness matrix, and then sum each element stiffness matrices to assemble the stiffness matrix for entire domain $\Omega$.

In this viewpoint, the FEM establishes an algebraic relationship between nodes. 
If the elements are chosen in a specific grid pattern, this algebraic relationship may
become equivalent to a finite difference stencil. 
The discretization of the Dirac equation in this paper is an extension
of previous works presented in~\cite{AndersonJCP2007,LimSURIO2013,Lim2014}.

There are several different choices for interpolation functions $n_i$.
For example, we use Lagrange tensor elements for the finite element spaces. In 1+1 case, the Lagrangian interpolation polynomials are:
\begin{eqnarray}
\label{eqn:lag_el1}
n_1(x,t)  = (h-x)(h-t)/h^2 \; ,\label{eqn:lag_el1} \\
n_2(x,t)  = x(h-t)/h \; ,\label{eqn:lag_el2} \\
n_3(x,t)  = (h-x)t/h \;, \label{eqn:lag_el3} \\
n_4(x,t)  = xt/h^2 \label{eqn:lag_el4} \; .
\end{eqnarray}
where $h$ is size of element

For 2+1 and 3+1 cases, we examine the interpolation polynomials with respect to each coordinates $x, y, z, t$. Different choice of interpolation function will provide different numerical efficiencies. 
A detailed study of interpolation function for 1+1 case is presented in~\cite{Vaselaar2014}. Detailed element stiffness matrix calculations are
provided in~\ref{appx:mat_elem}.

Due to the large size of the unsymmetrical system of the discretized matrix, iterative methods based on Krylov subspace (KSP) methods such as the generalized minimal residual method (GMRES)~\cite{SaadSIAM1986} are used. 

\section{Implementation Description}
\label{sec:software}

In this section, we describe our implementation details. We utilize
PETSc (Portable Extensible Toolkit for Scientific Computation) to build and solve our system.
PETSc, developed at Argonne National Laboratory, is at its core a highly efficient library for parallel linear algebra. It was designed for use in C, C++, Fortran, and Python. PETSc handles the parallel distribution of matrices and vectors, in addition to a variety of linear and nonlinear solvers that ease the use of parallel computing.  

Since the Dirac equation contains complex number, there are several important steps to import PETSc correctly. Here, we describe details on configuration and installation of PETSc 

To configure PETSc on a Linux or Mac OS to use our code, the following is required. First after downloading PETSc from~\cite{petsc-web-page}, there are two environmental variables that need to be defined. \texttt{PETSC\_DIR} is used to point to the directory where PETSc is located, for example \texttt{\$HOME\textbackslash user\textbackslash petsc\textbackslash} , while \texttt{PETSC\_ARCH} is the build name. \texttt{PETSC\_ARCH} does not need to be named anything specific but is commonly used to identify how PETSc was configured. Assuming Linux terminal an example of these would be, 
\begin{lstlisting}
 export PETSC_DIR=$HOME\user\petsc\   
 export PETSC_ARCH=complex_petsc  
\end{lstlisting}

Note that we highly recommended you to declare \texttt{PETSC\_DIR} and \texttt{PETSC\_ARCH} into your bash files such as \texttt{.bashrc} or \texttt{.bash\_profile} otherwise \texttt{PETSC\_DIR} and \texttt{PETSC\_ARCH} will need to be redefined every time you reopen the terminal. This can be done by adding above export commands into your bash files

Before PETSc can be configured, first make sure that you have compatible C, C++, Fortran, and python compliers installed; then run and follow the given instructions. 

\begin{lstlisting}
 ./configure --download-fblaslapack --download-mpich 
             --with-cc=<your C compiler> 
             --with-cxx=<your C++ compiler> 
             --with-fc=<your Fortran compiler>
             --with-scalar-type=complex
\end{lstlisting}

After successful configuration, PETSc introduce next step to install automatically. Or, you can simply type \texttt{make all check} to install PETSc. 

\subsection{PETSc configure options explained}

\texttt{fblaslapack} option installs both BLAS and LAPACK which are both numerical linear algebra libraries written in Fortran. If already installed this option can be disregarded, however if PETSc configure cannot find it in PATH run \texttt{ --with-blaslapack-dir=\textless location of BLAS/LAPACK\textgreater} .

\texttt{mpich} is an option for MPI communication, if a MPI is already installed configure without this command, however if it is not found a warning will appear during the configure process and the option \texttt{ -–with-mpi-dir=\textless location of MPI\textgreater} .

\texttt{\textless \_  Compiler\textgreater} is the name of whatever compiler you have installed for that language. 

\texttt{--with-scalar-type=complex} must be called for our code to work as default PETSc assumes real number system. 

More detailed information about the configuration and installation can be found in PETSc official website~\cite{petsc-web-page}.

\subsection{Dirac software Algorithm}
\begin{algorithm}
\caption{Main Algorithm}
\label{algo:dirac}
\begin{algorithmic}[1]
\State Given global desired mesh construct global size of vectors
\State Define initial conditions of $\psi$ 
\State Create Dirac $D_w$ matrix \Comment{\eqref{eqn:Dw}}
\State Separate $D_w$ into $D_{w22}$ and $D_{w21}$ \Comment{\eqref{eqn:dirac_weakIVP}} 
\State Create $A$ matrix using $D_{w22}$ and $b$ vector using $D_{w21}$ \Comment{Construct $Ax=b$}
\State Solve the linear system by utilizing KSP 
\end{algorithmic}
\end{algorithm}

Algorithm~\ref{algo:dirac} shows our main algorithm to solve the Dirac equation using spacetime FEM. 
We also describe parameters in this code in~\ref{appx:arti}

\section{Results}
\label{sec:results}
In this section, we provide several test cases to demonstrate
validations and functionalities. All tests are performed in LANL
supercomputer Badger and SDState Roaring Thunder. 
Badger is an 4-SU cluster running RHEL Linux v.7.7; it has dual socket 2.1 GHz 18 core Intel Broadwell E5 2695v4 processor with 45MB of cache and 128GB of RAM on each node.
Roaring Thunder consists of 56 compute nodes, 5 large memory nodes, 4 NVIDIA GPU nodes (V100/P100), and a 1.5 PB high-performance GPFS parallel file system. 

\subsection{Gaussian Plane Wave}
As a first step, we examine Gaussian plane wave solutions. 
Detailed derivation for plane wave solution is presented in~\ref{appx:gauss}.

\begin{figure*}[t!]
    \centering
    \begin{subfigure}[t]{0.5\textwidth}
        \centering
        \includegraphics[width=6cm]{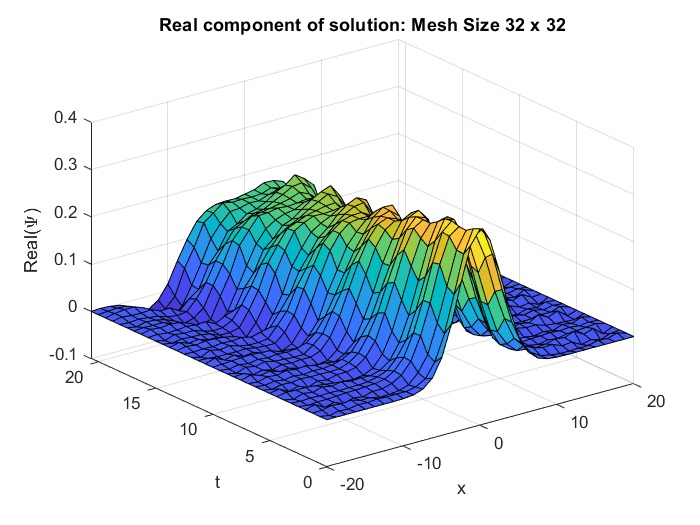}
        \caption{Real part of the solution}
    \end{subfigure}%
    ~ 
    \begin{subfigure}[t]{0.5\textwidth}
        \centering
        \includegraphics[width=6cm]{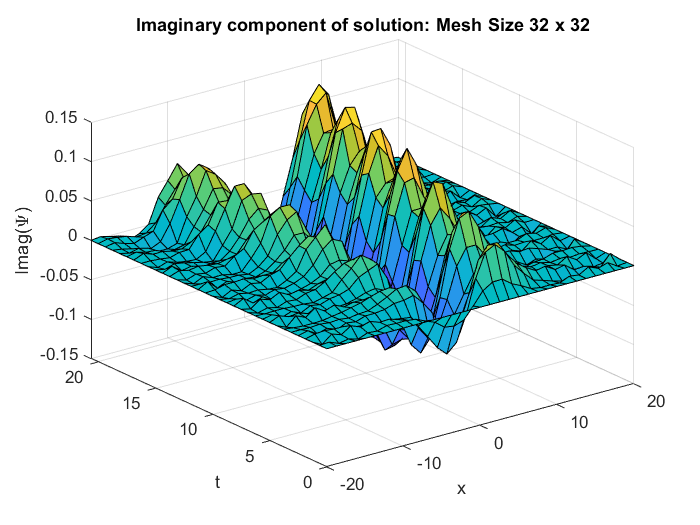}
        \caption{Imaginary part of solution}
    \end{subfigure}
    \caption{Gaussian plane wave solution for $1+1$ case. We use $32 \times 32$ meshes on the domain $-20 \leq x \leq 20$ and $0 \leq t \leq 20$ with a initial wave packet centered at $x=0$}
    \label{fig:gauss2d}
\end{figure*}

Fig.~\ref{fig:gauss2d} shows an example plane wave solution in $1+1$ case. We use $32 \times 32$ meshes on the domain $-20 \leq x \leq 20$ and $0 \leq t \leq 20$ with a initial wave packet centered at $x=0$. 

Since we can obtain the analytic expression for Gaussian wave packet, we compare our numerical results with analytic solution to examine our toolkit for all different dimensional cases.

For these testing, we varied mesh sizes, solvers, and initial amplitude by introducing additional scalar value to initial data with different dimensional cases with respect to each dimensions on the 
domain $0 \leq x \leq 1$ and $0 \leq t \leq 1$ with an initial wave packet centered at $x=0.5$. This particular choice of domain is small enough to resolve all wave motions with different mesh sizes. 
Our desired tolerance is $10^{-7}$ for all tests and we compute relative error by $|(\Psi_{\textrm{numeric}} - \Psi_{\textrm{exact}}) / \Psi_{\textrm{exact}}|\times 100$.

\begin{table}[]
\begin{tabular}{llll}
\hline
Mesh Size ($Nt \times Nx$) & Solver & Initial Amplitude & Relative Errors (\%) \\ \hline
$ 32 \times 32$ & GMRES  & 1 &  1.1332\\
$ 64 \times 64$ & GMRES  & 1 &  0.2863\\
$ 128 \times 64$ & GMRES & 1 &  0.1127\\ 
$ 32 \times 32$ & GMRES  & 0.5 &  1.2012\\
$ 64 \times 64$ & GMRES  & 0.5 &  0.0897\\
$ 128 \times 64$ & GMRES & 0.5 &  0.0503\\ 
$ 32 \times 32$ & BiCGSTAB  & 1 &  1.3119\\
$ 64 \times 64$ & BiCGSTAB  & 1 &  0.3245\\
$ 128 \times 64$ & BiCGSTAB & 1 &   0.1574\\ 
$ 32 \times 32$ & BiCGSTAB  & 0.5 &  1.2878\\
$ 64 \times 64$ & BiCGSTAB  & 0.5 &  0.2919\\
$ 128 \times 64$ & BiCGSTAB & 0.5 &  0.1073\\ 
\hline
\end{tabular}
\caption{1+1 case}
\label{table:plane:2d}
\end{table}

\begin{table}[]
\begin{tabular}{llll}
\hline
Mesh Size ($Nt \times Nx \times Ny$) & Solver & Initial Amplitude & Relative Errors (\%)\\ \hline
$ 24 \times 24 \times 24$ & GMRES  & 1 & 2.7147 \\
$ 32 \times 32 \times 32$ & GMRES  & 1 &  1.1526\\
$ 64 \times 32 \times 32$ & GMRES  & 1 &  0.5724\\ 
$ 24 \times 24 \times 24$ & GMRES  & 0.5 &  2.5958\\
$ 32 \times 32 \times 32$ & GMRES  & 0.5 &  1.0984\\
$ 64 \times 32 \times 32$ & GMRES  & 0.5 &   0.5322\\ 
$ 24 \times 24 \times 24$ & BiCGSTAB  & 1 & 2.7981 \\
$ 32 \times 32 \times 32$ & BiCGSTAB  & 1 &  1.1673\\
$ 64 \times 32 \times 32$ & BiCGSTAB  & 1 &  0.5802\\ 
$ 24 \times 24 \times 24$ & BiCGSTAB  & 0.5 &  2.7093\\
$ 32 \times 32 \times 32$ & BiCGSTAB  & 0.5 &  1.1424\\
$ 64 \times 32 \times 32$ & BiCGSTAB  & 0.5 &   0.5427\\ 
\hline
\end{tabular}
\caption{2+1 case}
\label{table:plane:3d}
\end{table}

\begin{table}[]
\begin{tabular}{llll}
\hline
Mesh Size ($Nt \times Nx \times Ny \times Nz$) & Solver & Initial Amplitude & Relative Errors (\%) \\ \hline
$ 12 \times 12 \times 12 \times 12$ & GMRES  & 1 & 5.3212 \\
$ 24 \times 24 \times 24 \times 24$ & GMRES  & 1 & 0.3426 \\
$ 32 \times 32 \times 32 \times 32$ & GMRES  & 1 & 0.1052 \\
$ 12 \times 12 \times 12 \times 12$ & GMRES  & 0.5 & 5.1447 \\
$ 24 \times 24 \times 24 \times 24$ & GMRES  & 0.5 & 0.3168 \\
$ 32 \times 32 \times 32 \times 32$ & GMRES  & 0.5 & 0.1136 \\
$ 12 \times 12 \times 12 \times 12$ & BiCGSTAB  & 1 &  5.2618\\
$ 24 \times 24 \times 24 \times 24$ & BiCGSTAB  & 1 &  0.3157\\
$ 32 \times 32 \times 32 \times 32$ & BiCGSTAB  & 1 &  0.1039\\
$ 12 \times 12 \times 12 \times 12$ & BiCGSTAB  & 0.5 &  5.5192\\
$ 24 \times 24 \times 24 \times 24$ & BiCGSTAB  & 0.5 &  0.3321\\
$ 32 \times 32 \times 32 \times 32$ & BiCGSTAB  & 0.5 &  0.1022\\
\hline
\end{tabular}
\caption{3+1 case}
\label{table:plane:4d}
\end{table}

Tables~\ref{table:plane:2d},~\ref{table:plane:3d},~\ref{table:plane:4d} show the Gaussian wave studies for 1+1, 2+1, and 3+1 respectively. We choose two different KSP solvers, GMRES and BiCGSTAB. As we increase number of meshes, relative errors decrease for all cases. 
We observe that both GMRES and BiCGSTAB agree with analytic solution well for all cases. Furthermore varying initial amplitude value didn't change solution quality too. 
For $3+1$ case, there is comparably larger error for smallest mesh size because mesh size for each direction might be small to resolve the system well. However, as we increase the mesh size, relative error decreases significantly. Note the matrix size for the smallest $3+1$ case is $82944 \times 82944$ which is reasonably large matrix system. 
In general, all cases show that our results agree well with analytic case.

The plane waves propagate vacuum so energy should be conserved. We monitor energy during the evolution to check energy loss. Using this energy loss,
we also measure order of convergence of our scheme.

\begin{figure}[ht]
  \centering
  \includegraphics[width=\textwidth]{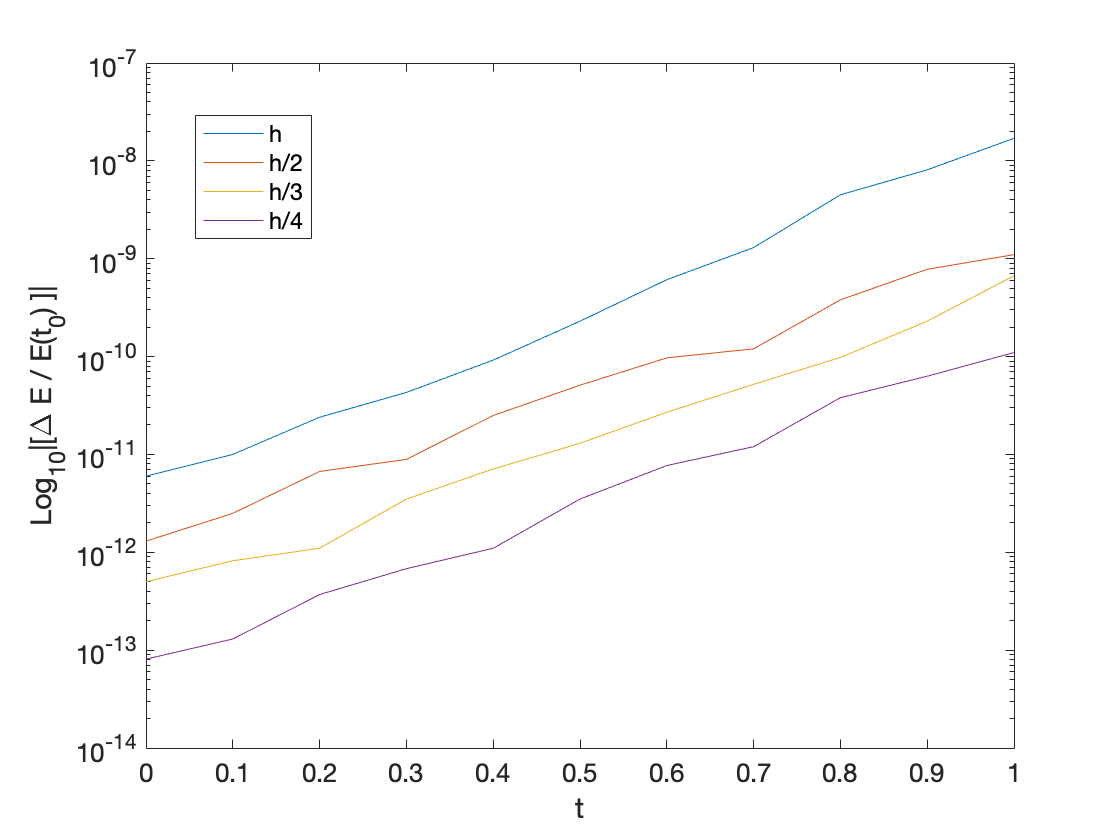}  
\caption{Energy loss plot for plane wave based on different resolutions. The mesh $h$ is a space resolution which can be obtained by dividing physical domain size by the number of collocation point. Here, $h$ is defined by $1/100 = 0.01$ which is chosen arbitrary. The energy loss $\Delta E(t) \equiv  E(t_f) - E(t_0)$ is plotted until the initial data reaches to end of our time domain $t_f = 1$. As resolution increases, energy loss decreases. Using this data, the self convergence test is evaluated $(||\Delta E_{h/4} - \Delta E_{h/2}||_2)/(||\Delta E_{h/2} - \Delta E_{h}||_2) = 4.091$ which indicates the order of self convergence is second order
}
\label{fig:Econv}
\end{figure}

Fig.~\ref{fig:Econv} shows the energy loss plots for plane wave solution. The energy loss is defined by $\Delta E(t) \equiv  E(t_f) - E(t_0)$ where $t_0$ is initial time and $t_f$ is final time i.e. end of our time domain. Ideally, the energy loss $|\Delta E(t)|$ should be zero. As shown in Fig.~\ref{fig:Econv}, energy loss decreases as resolution increase. Using this, a self convergence test is performed by evaluating $(||\Delta E_{h/4} - \Delta E_{h/2}||_2)/(||\Delta E_{h/2} - \Delta E_{h}||_2)$. This was taken where the same initial conditions were used but with different resolutions. The self convergence test value is 4.091 which indicates second order convergence.

\subsection{Zitterbewegung}
Zitterbewegung (`trembling motion' in German)~\cite{DiracQM1964} is a phenomena exclusive to relativistic quantum mechanics describing the oscillatory behavior of a wave packet which is intrinsic to the Dirac equation. This is quite different from the smooth relaxation of a packet predicted by the Schr\"{o}dinger equation. 

There are many different discussions on Zitterbewegung and its
interpretation. In the Newton-Wigner
theory~\cite{Newton1949}, the position operator leaves the
positive and negative energy sub-spaces invariant and thus it does
not display the Zitterbewegung behavior. In the Foldy-Wouthuysen
representation~\cite{Foldy1950}, this operator is identical to the standard
position operator. Regardless of these interpretation, 
we will focus on Zitterbewegung in position, $x(t)$ as a
test case for our implementation.

Using Heisenberg's picture, we can
obtain analytic expression on position operator
\begin{align}
i \dot{x} &= [H_0, x] ,\label{eqn:heisen:pos} 
\end{align}
where $H_0$ is free-field Dirac Hamiltonian. 
In this section, we keep $c$ to track physical constant.

Solving these equations will provide operator solutions
such that
\begin{align}
\label{eqn:ziiter:xt}
x(t) &= x + c^2 p_x H_0^{-1} t + \frac{ic}{2}(cp_x + H_0^{-2} 
- \alpha_x H_0^{-1}) (1-e^{-2iH_0t}) .
\end{align}
Note that the last term Eqn.~\ref{eqn:ziiter:xt} shows 
oscillatory behavior.

We are interested to obtain expectation value to
compare this analytic expression with our numerical
simulation.

To display this behavior using our method, the initial condition was set to a Gaussian wave packet centered at $x=0$.
\begin{eqnarray}
\psi(x,0) & = & \bigg(\frac{1}{32 \pi}\bigg)^{1/4} e^{-\frac{x^2}{16}} \begin{pmatrix} 1 \\ 1 \end{pmatrix}
\end{eqnarray}
The spatial expectation value $<x(t)>$ was calculated using the standard position operator. The following integrals were evaluated numerically using trapezoidal sums.
\begin{eqnarray}
<x(t)> & = & \int\limits_D x||\psi(x,t)||_2^2 dx .
\end{eqnarray}

\begin{figure}[ht]
  \centering
  \includegraphics[width=.9\linewidth]{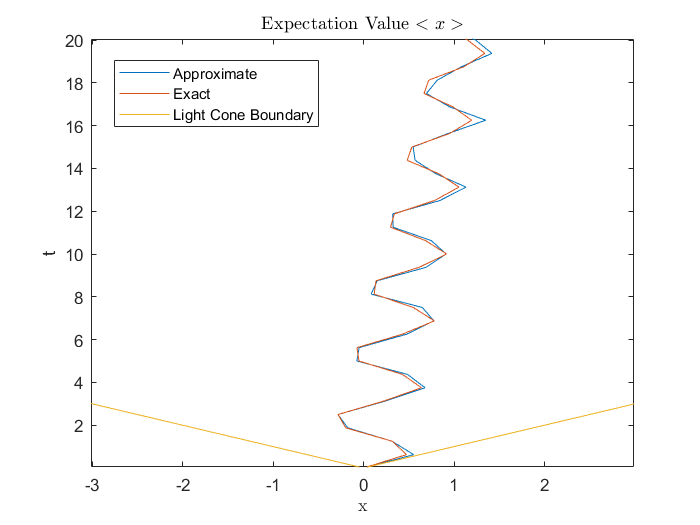}  
\caption{Zitterbewegung behavior).
}
\label{fig:zitter}
\end{figure}

Figure~\ref{fig:zitter} shows the numerical and exact spatial expectation value, $<x(t)>$
as well as the light cone boundary. The light cone boundary represents the trajectory of a 
particle at the speed of light. Any results outside this boundary would be non-physical,
as it would imply part of the wave packet is traveling faster than the speed of light.
As it is presented in Fig.~\ref{fig:zitter}, the numerical results seem to compare well to the
analytic solution and lie inside the light cone boundary as expected.

\subsection{Klein Paradox}
The Klein paradox~\cite{Klein1929} is a well-known example for which the single particle
interpretation of the Dirac equation can lead to some paradoxical predictions. 
In non-relativistic quantum mechanics, the wave function in the region where potential, $V$,
is non-zero is decaying exponentially if the energy, $E$, of the incident wave is lower than the $V$. 
Thus, most of the wave function is reflected and transmission coefficient is negligible. 
However, in the relativistic case, a new phenomenon appears when $E < V- mc^2$. In this regime, a plane wave solution can exist, resulting in a non-negligible transmission coefficient even if $E<V$. 
Many previous works have been reported to``resolve" the Klein paradox in the context of the Dirac see picture
and second quantization~\cite{Kerkora2004, Dombey1999}
In this subsection, we explore this phenomena in 1+1 case.

The Dirac equation for a particle in the presence of an external scalar potential field is
\begin{eqnarray}
\begin{pmatrix} i \partial_t + V - m & i \partial_x \\ -i \partial_x & -i \partial_t - V - m \end{pmatrix} \begin{pmatrix}
\psi^1 (x) \\
\psi^2 (x) 
\end{pmatrix} & = & 0.
\end{eqnarray}

Gaussian wave packet is chose for initial data such that
\begin{eqnarray}
\label{eqn:klein:id}
\psi(x,t=0) =  \bigg(\frac{1}{32 \pi}\bigg)^{1/4} e^{-100(x-0.6)^2}e^{35ix}\begin{pmatrix} 1 \\ 1 \end{pmatrix} .
\end{eqnarray}
This is a wave packet centered at $x = 0.6$ with an average momentum of $35$. Also, we consider a smooth potential to avoid numerical
issues related with discontinuous function. 
This potential, called Sommerfeld potential
is given by
\begin{eqnarray}
\label{eqn:sommer_pot}
V(x) & = & \frac{V_0}{2}[1 + \tanh((x-x_0)/L)].
\end{eqnarray}
Here, $V_0$ and $x_0$ are the magnitude and central location of the potential step, respectively, and $L$ corresponds to the steepness of the transition to the step.

\begin{figure}[ht]
  \centering
  \includegraphics[width=\textwidth]{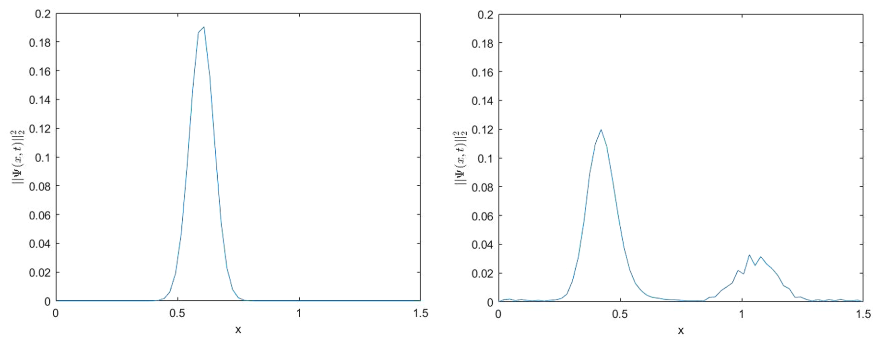}  
\caption{Example wave density $||\psi(x,t)||^2_2$ before (left) and after (right) the interaction of the 
initial wave packet with the potential barrier. Potential barrier is located at $x_0=0.8$.
}
\label{fig:klein_wave}
\end{figure}

Fig.~\ref{fig:klein_wave} shows the wave distribution $||\psi(x,t)||_2^2$ before and after scattering at the Sommerfeld potential. In this figure, the parameters for this potential were set to $V_0 = 36.1$, $L = 0.001$, and $x_0 = 0.8$,

The potential (Eqn.~\ref{eqn:sommer_pot}) is already studied by~\cite{Sommerfeld1921, Sauter1931}.
The exact formula for the transmission coefficient for a Sommerfeld potential is given by
\begin{eqnarray}
\label{eqn:klein:T}
T = \frac{\sinh(\pi p L)\sinh(\pi p' L)}{\sinh(\pi(V_0 + p + p')\frac{L}{2})\sinh(\pi(V_0 - p - p')\frac{L}{2}},
\end{eqnarray}
where $p = \sqrt{(E_p - V_0)^2 - 1}$, \ $p' = -\sqrt{E_p^2 - 1}$, and $E_p = \sqrt{p_0^2 + 1}$. Using this, reflection coefficients can be obtained by $R=1-T$.

To compare this result with an analytic expression, the $T$ and $R$ were calculated
such that
\begin{eqnarray}
\label{eqn:klein:coeffs}
T = \frac{\int_{\Omega^+} ||\psi(x,t_f)||_2^2 dx}{\int_\Omega ||\psi(x,t_f)||_2^2 dx}, \;\;
R =\frac{\int_{\Omega^-} ||\psi(x,t_f)||_2^2 dx}{\int_\Omega ||\psi(x,t_f)||_2^2 dx} ,
\end{eqnarray}
where $\Omega$ is whole spatial domain, $\Omega^+$ is transmitted domain, and $\Omega^-$ is reflected domain with $t_f$ being the final time value, after the wave has scattered at the potential. These integrals were evaluated numerically using trapezoidal sums.

\begin{figure}[ht]
  \centering
  \includegraphics[width=.9\linewidth]{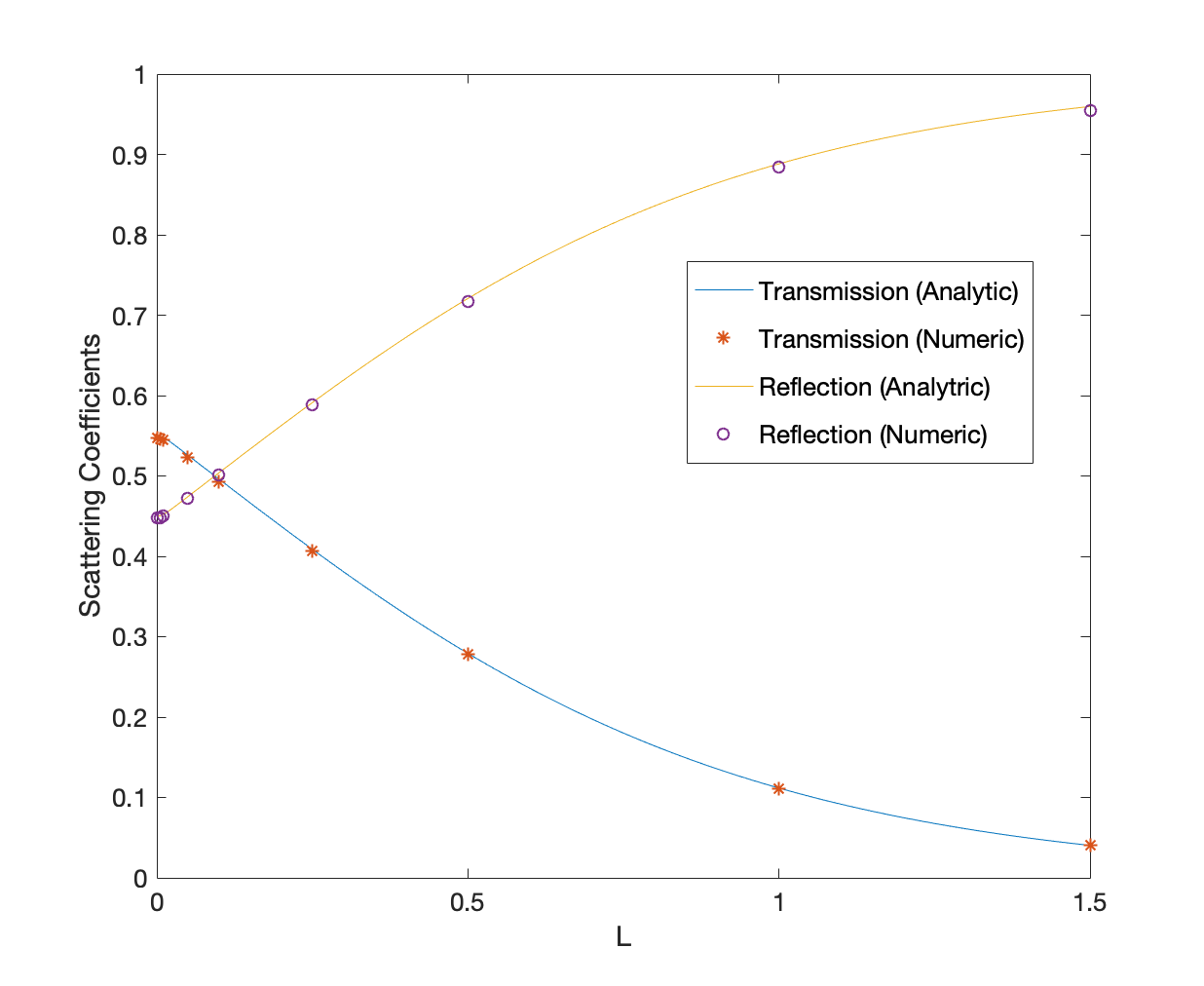}  
\caption{Scattering coefficients as function of the steepness of the transition to the step. We vary different values of L= [0.001,0.005,0.01,0.05,0.1,0.25,0.5,1.0,1.5] to compare with analytic expression (Eqn~\ref{eqn:klein:T}).
}
\label{fig:klein_scattering}
\end{figure}

Fig.~\ref{fig:klein_scattering} shows scattering coefficients as a function of the steepness of the transition to the step, $L$.
For this test, we use $256 \times 256$ mesh size and after $10,000$ iterations of the GMRES.
Different $L$ were tested to compare with analytic values of transmission and reflection coefficients. All relative errors are less than 0.5\% which shows our numerical results agree well with analytic expression. 

\subsection{Performance Tests}

We perform both weak and strong scaling tests to measure parallel efficiency of our implementation. 
on the LANL supercomputing cluster, Badger. 

The plane wave with different dimensions was used for both tests. For jobs with number of
MPI ranks from 1 to 32, we used a single node, and for larger jobs, we use multiple nodes with
32 rank per nodes. For strong scaling tests, we use meshes of $256 \times 256$, $64 \times 64 \times 64$, and $32 \times 32 \times 32 \times 32$ elements for $1 + 1$, and $2 + 1$, and $3 + 1$ respectively. 
For weak scaling we use 256, 192, and 128 elements per rank for $1 + 1$, and $2 + 1$, and $3 + 1$ respectively. All tests are evolved with 100 GMRES iterations. Then we compute strong scaling efficiency, $t_1/(N \times t_N) \times 100\%$ and weak scaling efficiency, $t_1 / t^{'}_N \times 100\%$ where $t_1$ is the amount of time to complete a task with 1 processing element, $t_N$ is the amount of time to complete the same task with $N$ processing elements, and $t^{'}_N$ is the amount of the time to complete $N$ of the same task with $N$ processing elements. 

\begin{figure}[ht!]
\centering
\begin{tabular}{cc}
\includegraphics[width=.49\textwidth]{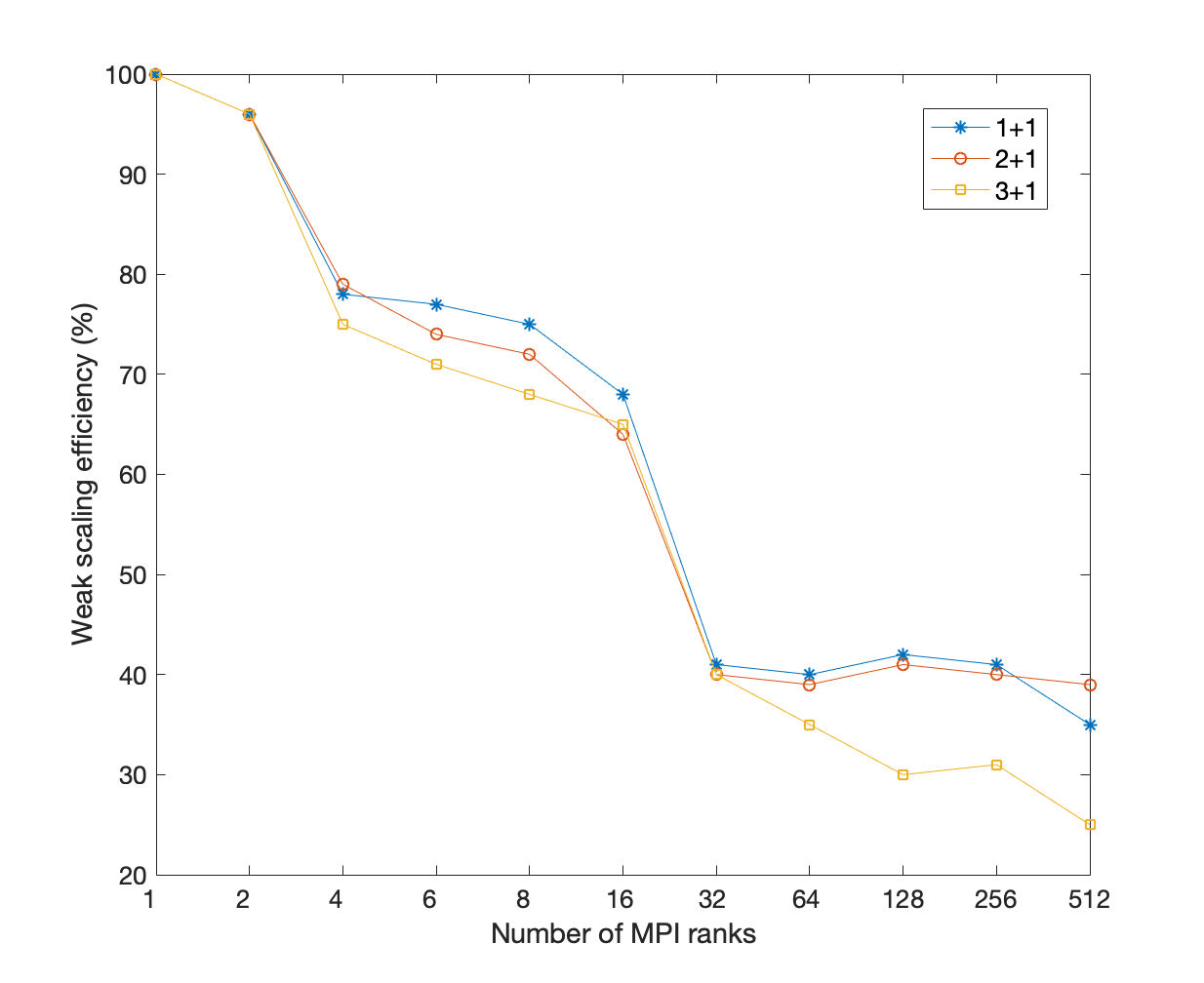} &
\includegraphics[width=.49\textwidth]{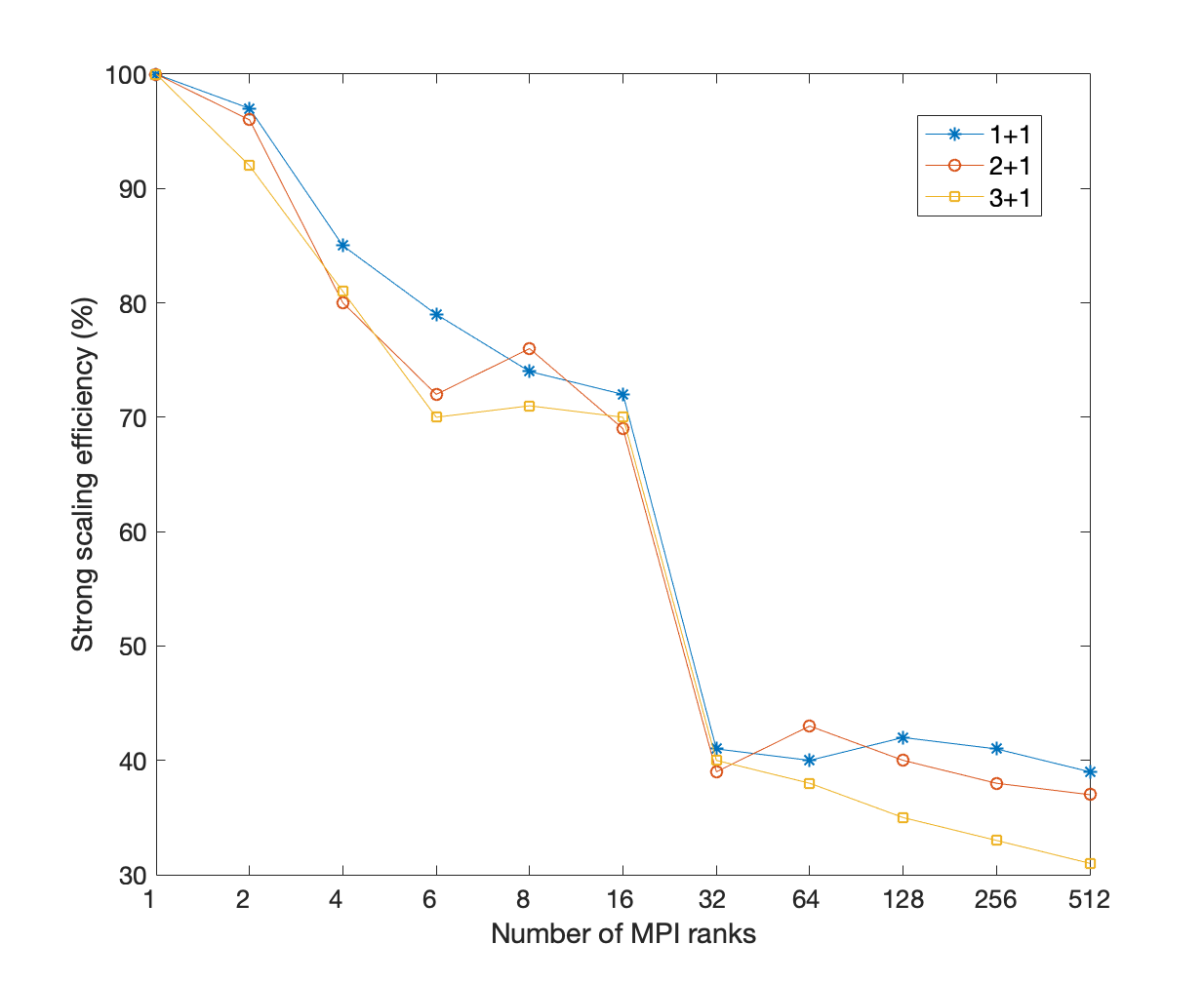}
\end{tabular}
\caption{Weak (left) and strong (right) scaling efficiency on Badger clusters. 
Strong scaling tests used meshes of $512 \times 512$, $128 \times 128 \times 128$, and $64 \times 64 \times 64 \times 64$ elements for $1 + 1$, and $2 + 1$, and $3 + 1$ respectively. 
Weak scaling used 256, 192, and 128 elements per rank for $1 + 1$, and $2 + 1$, and $3 + 1$ respectively. The plane wave functions are evolved with 100 GMRES iterations. } 
\label{fig:scaling}
\end{figure}

Fig.~\ref{fig:scaling} shows strong scaling and weak scaling efficiencies for $1 + 1$, and $2 + 1$, and $3 + 1$ cases. Note that efficiencies degrade between 32 and 64 ranks due to the inter-node data transfer, but then remain relatively flat.
Overall all cases show comparable scaling. Having a better treatment to partitioning matrix system for higher numbers of MPI ranks will provide better scaling efficiencies.

\section{Conclusion and Future Works}
\label{sec:conclusion}

In this work, we present a spacetime FEM
to solve the Dirac equation. We demonstrate several different
application examples, Gaussian plane waves, Zittwerbewegung, and Klein paradox to validate our method and implementation. All of these tests show good agreement with analytic cases. Further, we explore our parallel efficiency. Although certain limitations need to overcome such as inter-node data transfer, all cases show comparable scaling. 

As its nearest goal, this method will solve more realistic problems
including an inhomogeneous Dirac equation.
Further, complicated and realistic problems will require
more computational costs thus exploring a proper preconditioner
for the system to solve the problem is required. In~\cite{AndersonJCP2007,Lim2014}, they applied time-additive
Schwarz method as a time decomposition method.
Since the nature of Dirac equation is complex, it will be interesting
subject to apply the time decomposition idea for this problem and monitor how linear solver such as GMRES performance will be changed. This might provide better performance.


\section*{Acknowledgement}

HL is supported by the LANL ASC Program and LDRD grants 20190021DR. This work used resources provided by the LANL Institutional Computing Program. LANL is operated by Triad National Security, LLC, for the National Nuclear Security Administration of the U.S.DOE (Contract No. 89233218CNA000001). This article is cleared for unlimited release LA-UR-21-22066.

\appendix
\section{Matrix Element Calculations}
\label{appx:mat_elem}
We consider 1+1 case in Eqn.~\ref{eqn:diracweakIVP} for simplicity.
Then equation becomes
\begin{align}
\label{diracWeak1p1}
D_{ij} = \int_\Omega \notag n_j(x,t) 
         \bigg(i \hbar 
         \begin{pmatrix} 
         1 & 0 \\ 
         0 & -1 
         \end{pmatrix}
         \partial_t 
         + i \hbar 
         \begin{pmatrix} 
         0 & 1 \\ 
         -1 & 0 
         \end{pmatrix}
         \partial_x 
         - m \mathbb{I} \bigg)n_i(x,t) dxdt
\end{align}
Note that $D_{ij}$ is written as a double integral of three separate terms, which can be separated as
\begin{eqnarray}
D_{ij} = \begin{pmatrix} 1 & 0 \\ 0 & -1 \end{pmatrix} i \hbar  D_{ij}^t + i \hbar \begin{pmatrix} 0 & 1 \\ -1 & 0 \end{pmatrix} D_{ij}^x - m \mathbb{I} D_{ij}^0, \notag
\end{eqnarray}
where the individual terms $D_{ij}^t$, $D_{ij}^x$, and $D_{ij}^0$ are given by
\begin{eqnarray}
D_{ij}^t & = & \int_0^h \int_0^h n_j(x,t) \partial_t n_i(x,t) dxdt \notag \\
D_{ij}^x & = & \int_0^h \int_0^h n_j(x,t) \partial_x n_i(x,t) dxdt \notag \\
D_{ij}^0 & = & \int_0^h \int_0^h n_j(x,t) n_i(x,t) dxdt. \notag
\end{eqnarray}
Using Eqns.~\ref{eqn:lag_el1}~-~\ref{eqn:lag_el4}, we can compute all of these integrations. As results, each $D^x_{ij}$, $D^y_{ij}$, $D^z_{ij}$ are $4 \times 4$ matrices. We can extend this idea to higher dimensional cases. We increase spatial dimension into $n_i$, add extra spatial derivatives for $D_{ij}$, and increase dimension of gamma matrices for each higher dimension respectively. Thus, we have four $8 \times 8$ matrices and five $16 \times 16$ matrices for $2+1$ and $3+1$ respectively.
We use \texttt{MAXIMA}~\cite{maxima} to perform integration.
Our \texttt{Maxima} script can be found in~\url{https://gitlab.com/resundermann/dirac/-/tree/master/tools/integration}

\section{Solutions for Plane Gaussian Waves}
\label{appx:gauss}
The first case of the Dirac equation examined in this study is the free-field form, meaning that the particle described by the Dirac equation has no external fields or forces interacting with it. The exact solution to the free-field Dirac equation is relatively simple given the initial state $\psi(x,0)$, making it an appropriate candidate for comparison to a numerical solution. Suppose the initial state of the free particle is a wave packet given by 
\begin{eqnarray}
\psi(x,0) & = & \bigg(\frac{1}{32 \pi}\bigg)^{1/4} e^{-\frac{x^2}{16} + i\frac{3x}{4}} \begin{pmatrix} 1 \\ 1 \end{pmatrix}. \notag
\end{eqnarray}
A wave packet is a superposition of waves with a range of momentum which is often used to describe a particle in quantum mechanics. Taking a Fourier transform of the initial state yields the wavefunction as a function of momentum rather than space, $\phi(p,t=0)$. Therefore, the formula for the wavefunction $\phi(p,t=0)$ is
\begin{eqnarray}
\phi(p,t=0) & = & \frac{1}{\sqrt{2 \pi}} \int\limits_{-\infty}^{\infty} \psi(x,0) e^{-ipx}dx \notag \\
& = & \frac{1}{\sqrt{2 \pi}} \int\limits_{-\infty}^{\infty} \bigg(\frac{1}{32 \pi}\bigg)^{1/4} e^{-\frac{x^2}{16} + i\frac{3x}{4}} \begin{pmatrix} 1 \\ 1 \end{pmatrix} e^{-ipx}dx. \notag
\end{eqnarray}
The above integral can be evaluated analytically and is shown in Appendix A. The result is the following wavefunction,
\begin{eqnarray}
\phi(p,t=0) & = & \bigg(\frac{8}{\pi} \bigg)^{1/4} e^{-4(p - \frac{3}{4})^2} \begin{pmatrix} 1 \\ 1 \end{pmatrix}. \notag
\end{eqnarray}
Now, this momentum-dependent wavefunction can be written as the linear combination of vectors $u_{pos}, u_{neg}$ so that the proper time-dependency can be assigned to each component [6]. This form of the wavefunction is
\begin{eqnarray} 
\phi(p,0) & = & \phi^+(p)u_{pos}(p) + \phi^-(p)u_{neg}(p). \notag
\end{eqnarray}
Here, the coefficients $\phi^{\pm}$ are the scalar products between the initial wavefunction and the vectors $u_{pos}$ and $u_{min}$, given by
\begin{eqnarray}
\phi^{\pm} & = & \langle u_{pos, neg}(p),\phi(p) \rangle _2, \notag \\
u_{pos}(p) & = & \frac{1}{\sqrt{2}} \begin{pmatrix} (1 + 1/\sqrt{k^2 + 1})^{1/2} \\ sgn(p)(1 - 1/\sqrt{k^2 + 1})^{1/2} \end{pmatrix}, \notag 
\end{eqnarray}
and
\begin{eqnarray}
u_{neg}(p) & = & \frac{1}{\sqrt{2}} \begin{pmatrix} -sgn(p)(1 - 1/\sqrt{k^2 + 1})^{1/2} \\ (1 + 1/\sqrt{k^2 + 1})^{1/2} \end{pmatrix}. \notag
\end{eqnarray}
Multiplying the upper and lower components by their respective time dependency, $e^{-iEt}$ and $e^{iEt}$ where $E = \sqrt{p^2 + m^2}$ (relativistic energy), yields the time dependent wavefunction, 
\begin{eqnarray}
\phi(p,t) & = & \phi^+(p)u_{pos}(p)e^{-iEt} + \phi^-(p)u_{neg}(p)e^{iEt}. \notag
\end{eqnarray}
Finally, taking the inverse Fourier transform of the wavefunction in momentum space yields the time dependent wavefunction in position space, written as
\begin{eqnarray}
\psi(x,t) & = & \int\limits_{-\infty}^{\infty}e^{ipx}(\hat{\phi}^+(p)u_{pos}(p,t) + \hat{\phi}^-(p)u_{neg}(p,t))dp. \notag 
\end{eqnarray}
This method can be extended to higher dimensional cases by adding extra spatial dimension.
\section{Artifact Description}
\label{appx:arti}

We maintain all our implementation using \texttt{git} version control system which is available in \url{https://gitlab.com/resundermann/dirac} as an open source. We use \texttt{Matlab} to analyze our data and plot the result. Description of our repository is listed in below
\begin{itemize}
    \item  {applications:} This holds the main driver functions for the 2D, 3D, and 4D codes.
    \item  {include:} Folder for the header files of each main and src. 
    \item  {miscell:} This is our miscellaneous folder, it contains documentation, old MATLAB code references and scripts we used in our testing. This is currently re-organizing.
    \item {src:} This folder contains the functions called in the main functions, including the Dirac functions that build the Dirac matrix, in addition to the functions that separate it into smaller matrices, handled in the Dirac and ShorteMat scripts respectfully.
    \item {tools:}  currently holds maxima files used in integration for the 2D, 3D, and 4D Dirac codes.
\end{itemize}
Our repository provide self-contained document for building the code and development work flow. 

The code has the functionality to change the number of elements in each dimension at run time. These are the x, y, z, t and angle parameters. Space dimensions default to 4, time 3, angle 30. To change any of these parameters use -variable value. For example, to change x to 10 when running the 4D code the option is “./RotationDirac4D -x 10”. 


\bibliography{refs}

\end{document}